\documentclass[aps,prc,twocolumn,superscriptaddress,floatfix,longbibliography,10pt]{revtex4-2}

\usepackage{graphicx}
\usepackage{hyperref}
\usepackage{float} 
\usepackage{xcolor}
\usepackage{amsmath}
\usepackage{amssymb} 
\usepackage[T1]{fontenc}
\usepackage[english]{babel}
\usepackage{subcaption}
\usepackage{textcomp,gensymb}
\usepackage{pifont} 
\usepackage{orcidlink} 

\usepackage{caption}

\makeatletter
\let\ORIbbl@fixname\bbl@fixname
\def\bbl@fixname#1{%
  \@ifundefined{languagealias@\expandafter\string#1}
    {\ORIbbl@fixname#1}
    {\edef\languagename{\@nameuse{languagealias@#1}}}%
}
\newcommand{\definelanguagealias}[2]{%
  \@namedef{languagealias@#1}{#2}%
}
\makeatother
\definelanguagealias{en}{english}

\begin{document}

\title{New Physics Search with the Optical Dump Concept at Future Colliders}

\author{Ivo Schulthess\orcidlink{0000-0002-5621-2462}}
\email[Corresponding author: ]{ivo.schulthess@desy.de}

\author{Federico Meloni\orcidlink{0000-0001-7075-2214}}

\affiliation{Deutsches Elektronen-Synchrotron DESY, Hamburg, Germany}



\date{\today}

\begin{abstract}
    Beam-dump experiments offer an opportunity to search for new physics beyond the Standard Model of particle physics. In this work, we explore the use of a high-energy photon beam on a fixed target. Such photons can be produced via Compton backscattering when colliding high-energy electrons with a high-intensity laser pulse, such as in setups of strong-field quantum electrodynamics experiments. We present various options for implementing such photon sources at future facilities and discuss how the photons can be used to search for axion-like particles. We compare the projected sensitivities of the proposed options with existing constraints, projections from current experiments, and direct searches at future colliders.
\end{abstract}

\maketitle

\section{Introduction}\label{sec:intro}

Although the Standard Model of particle physics has achieved remarkable success, it does not address several unresolved questions. They include, e.g., the observed dark matter and baryon asymmetry in our universe \cite{boucenna_theories_2014}. Many new theoretical models that try to answer these questions require new particles and forces. Their existence has to be probed experimentally. So far no particles beyond the Standard Model have been discovered~\cite{berlin_new_2024, navas_review_2024}. In case they existed, they could be difficult to detect because of their weak interaction with Standard-Model particles. 

One category is new light spin-0 particles, which can be scalar or pseudo-scalar. The latter is known as axions or axion-like particles (ALPs). In the mass range accessible by fixed-target experiments at future colliders, they could serve as a portal to the dark sector or as an explanation for the anomalous magnetic moment of the muon~\cite{dolan_revised_2017, curtin_long-lived_2019}.

\begin{figure}
    \centering
    \includegraphics[width=1.0\linewidth]{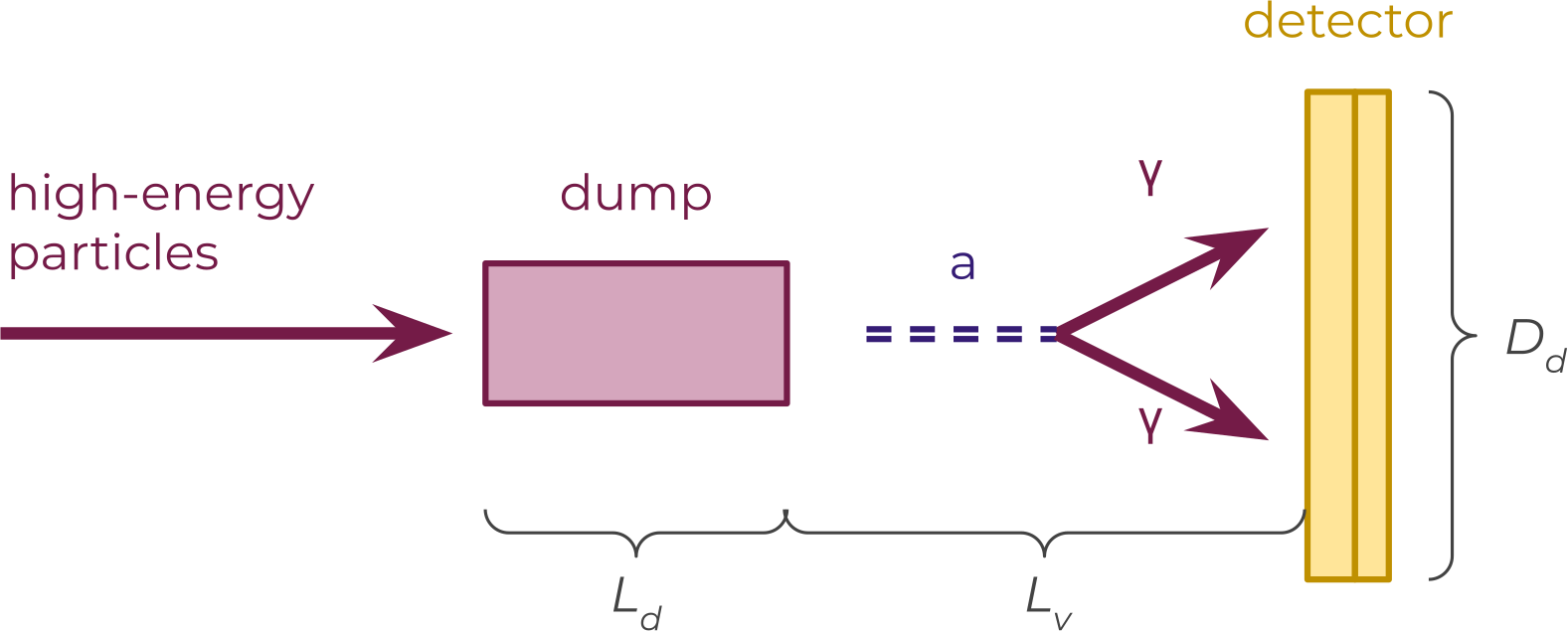}
    \caption{\it Schematic of the experimental configuration of a fixed-target or beam-dump experiment. The relevant geometrical parameters are the length of the dump $L_d$, the length of the decay volume $L_v$, and the detector diameter $D_d$. }
    \label{fig:schematic}
\end{figure}

To search for new physics in a fixed-target experiment, high-energy particles are shot on a usually high-Z target material. The production process we consider is the Primakoff production~\cite{primakoff_photo-production_1951}: photons interact with the nuclear electric fields of the target material, thereby producing ALPs. So far, this has been done mostly with charged particles such as protons, electrons, or positrons. For these, it is a two-step process: the charged particles produce Bremsstrahlung photons in the dump, which then can undergo Primakoff production. In addition to charged particles, high-energy photons can be used. In this case, the Bremsstrahlung process is evaded, and the photons can directly undergo the Primakoff production in the target material. The produced ALPs could decay into two photons on the way to the detector. These two photons could be detected if the decay occurs in the decay volume. The concept is schematically depicted in Fig.~\ref{fig:schematic}.

One way of producing a high flux of photons with GeV energy and above is by colliding a high-energy electron beam with a high-intensity laser. The interaction in this non-linear regime is described by strong-field quantum electrodynamics (QED). The parameter space of strong-field QED can be defined by two parameters. The first is the laser intensity parameter ${a_0 = e E_l \, / \, \omega_l m_e c}$, where $e$ is the elementary charge, $E_l$ the electric field amplitude of the laser, $\omega_l$ is the angular optical frequency of the laser, $m_e$ the electron mass, and $c$ the speed of light in vacuum. It can also be written in a Lorentz invariant way ${a_0 = \left( e^2 \left< (-p \cdot \mathcal{F})^2 \right> \, / \, \left[m_e k \cdot p \right]^2 \right)^{1/2}}$, where $\mathcal{F}$ is the classical field strength tensor, $p$ is the momentum of the electron, $k$ is the wavevector of the laser field, and ${\left< \cdot \right>}$ indicates the phase cycle average. The second parameter is the energy parameter $\eta = \gamma \frac{\hbar \omega_l}{m_e c^2} \left( 1 + \cos(\theta) \right)$ or $\eta = \hbar k \cdot p / m_e^2 c^4$, where $\gamma$ the Lorentz boost factor of the electron, $\theta$ the collision angle between the electron and the laser, and $\hbar$ the reduced Planck constant. The two main effects that can be observed in the electron-laser collision are the non-linear Compton process~\cite{compton_quantum_1923} and the non-linear Breit-Wheeler pair production~\cite{breit_collision_1934}. A recent review of the topic can be found in Ref.~\cite{fedotov_advances_2023}. 

In particular, the non-linear Compton process can be exploited to generate a high-intensity photon beam: For electrons, the laser pulse acts as a thick target. When they collide, they radiate their energy in the form of GeV photons. On the other hand, the produced photons are effectively free-streaming since the laser pulse is transparent to them. They can be used further downstream in the fixed-target or beam-dump experiment. This concept is called \textit{new physics search with the optical dump} or NPOD~\cite{bai_new_2022}. To describe this more formal, the following condition has to be fulfilled: 
\begin{equation}\label{eq:opticalDumpCondition}
    1/\omega_l \ll \tau_\gamma \lesssim t_l \ll \tau_{ee} \ ,
\end{equation}
where ${1/\omega_l \approx 0.4~\mathrm{fs}}$ is the time of a laser oscillation, ${\tau_\gamma \approx 10~\mathrm{fs}}$ is the time scale of the radiation process, ${t_l \approx 10 - 200~\mathrm{fs}}$ is the laser pulse length, and ${\tau_{ee} \approx 10^4 - 10^6~\mathrm{fs}}$ is the time scale of the pair-production process. The times are given for a laser intensity of ${a_0 \approx 3.4}$~\cite{bai_new_2022}. The condition Eq.~(\ref{eq:opticalDumpCondition}) means, that the interaction of the electrons with the laser pulse is long enough to radiate but short enough to not produce any electron-positron pairs. 

The NPOD concept was initially studied in the framework of LUXE which is a planned experiment at DESY in Hamburg (DE). It will collide electrons from the European XFEL with a high-intensity laser pulse~\cite{abramowicz_conceptual_2021, luxe_collaboration_technical_2024}. The main goal is to precisely test the transition from perturbative to non-perturbative QED. It is relevant to our fundamental understanding of light-matter and light-light interactions. It finds application in astrophysics, plasma physics, the interaction of charged particles with crystals, and future high-energy lepton colliders. The LUXE-NPOD study showed that it could test parts of the phase-space of the ALP-photon coupling, which is yet uncovered by other laboratory experiments~\cite{bai_new_2022}. The idea of NPOD was also conceptually studied for the International Linear Collider ILC with the same laser intensity as for LUXE (${a_0 \approx 3.4}$), called ILC-NPOD. It would extend the phase-space coverage to higher masses and coupling constants~\cite{aryshev_international_2023, soreq_probing_2021, irles_probing_2023}. 

In this work, we follow up on the ILC-NPOD study~\cite{aryshev_international_2023} and examine in more detail what can be done at future collider facilities. In Sec.~\ref{sec:setup}, we first explain the settings of the simulations. This includes the Primakoff production cross section, the geometrical settings, as well as the various photon sources considered, and their energy spectra. We then elaborate on the assumption that such an experiment can be considered background-free in Sec.~\ref{sec:background}. In Sec.~\ref{sec:results} we show the estimated coverage of the phase space and conclude in Sec.~\ref{sec:conclusion} with some aspects that should be considered in future studies.

\section{Setup}\label{sec:setup}

\subsection{Primakoff Cross Section}

In this study, we only consider the photoproduction of axion-like particles ALPs via the Primakoff production and the subsequent decay into two photons. This simple model requires only the coupling of ALPs to photons. It can also be applied similarly to scalar particles and other couplings, such as a fermionic coupling to electrons. 

\begin{figure}
    \centering
    \includegraphics[width=1.0\linewidth]{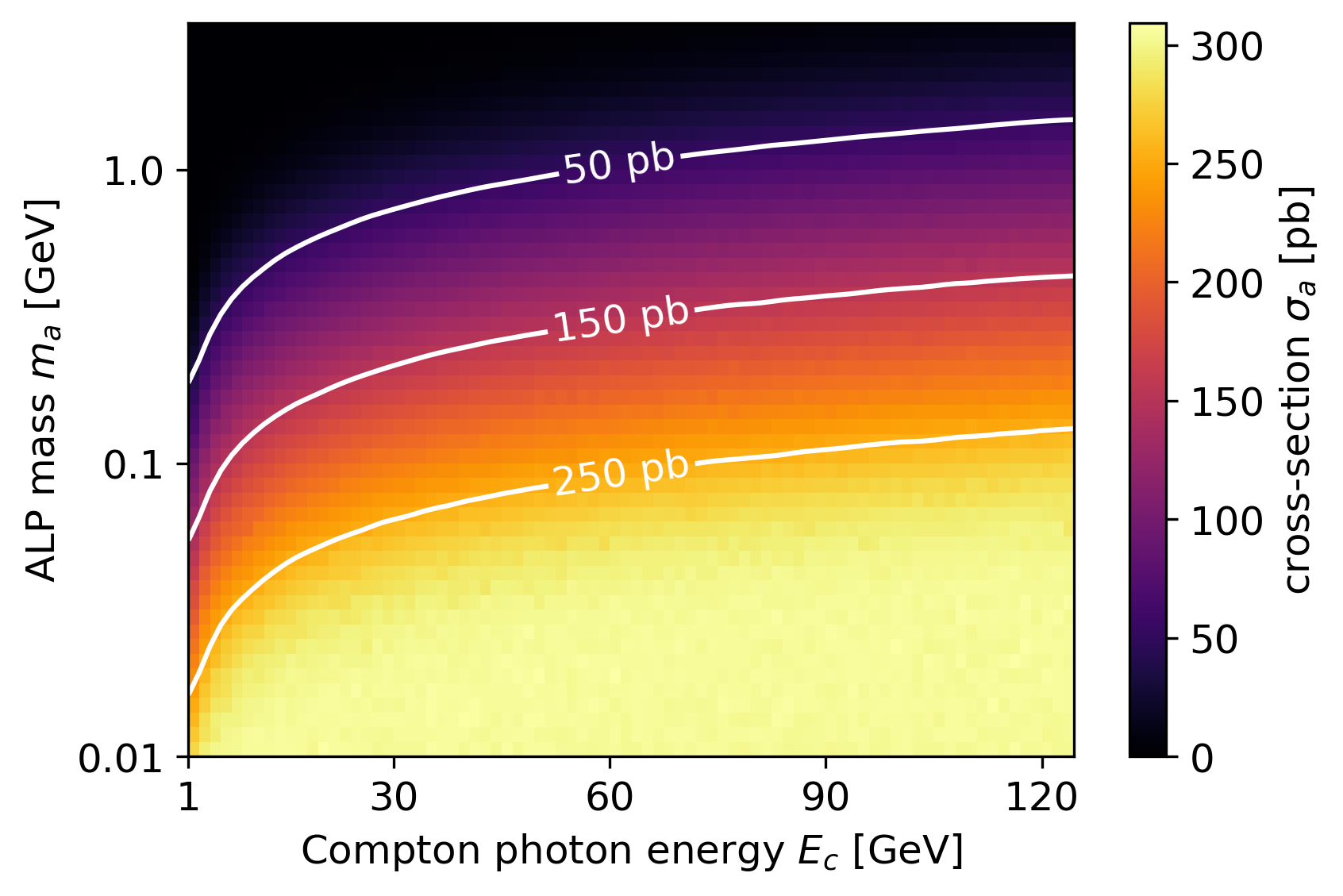}
    \caption{\it Results of the cross-section calculations for various Compton photon energies $E_c$ and ALP masses $m_a$. The calculated mechanism is the Primakoff production on the tungsten nuclei of the dump. }
    \label{fig:crossSection}
\end{figure}

We calculated the Primakoff production cross section and simulated the ALP signal distribution with the Monte Carlo simulation tool \textsc{Madgraph5 v2.8.1}~\cite{stelzer_automatic_1994, alwall_madgraph_2011, alwall_automated_2014, degrande_ufo_2012}. As in Ref.~\cite{bai_new_2022}, we use the UFO model~\cite{degrande_ufo_2012} from Ref.~\cite{brivio_alps_2017} and follow Refs.~\cite{chen_muon_2017, tsai_pair_1974, bjorken_new_2009} for the form factors. This was done for tungsten as a target material as it leads to the highest ALP production rates. The results of the cross-section calculations are presented in Fig.~\ref{fig:crossSection}. It shows that the production cross section is only weakly dependent on the incident photon energy, especially above 20~GeV. 

\subsection{Geometry}\label{sec:geometry}

The idealized geometry of the simulation is shown in Fig.~\ref{fig:schematic}. The high-energy photons undergo the Primakoff process at the front face of the dump on axis. The dump has a length $L_d$. The produced ALPs are only further considered if they decay in the decay volume of length $L_v$ after the dump but before the detector. The two decay photons are then further propagated and counted if both are within the detector acceptance of $D_d$ and each have a minimum energy of $E_\gamma \geq 0.5~\mathrm{GeV}$. 

Two different geometries were considered, summarized in Tab.~\ref{tab:simulationGeometry}: a short dump and decay volume of a total length of only 12~m (referred to as Zugspitze~\footnote{As the highest mountain of the Alps in Germany, Zugspitze serves as the name for the benchmark model}) allows access to shorter decay lengths and shorter proper lifetimes. This is generally less covered by existing constraints and therefore more interesting~~\cite{ohare_cajohareaxionlimits_2020}. However, a shorter dump can result in a significantly higher background since more particles can penetrate through it and reach the detector. A length of roughly $L_d = 2~\mathrm{m}$ is the minimum length required for the setup to still be in a background-free environment. More detailed background estimates are given in Sec.~\ref{sec:background}. The second geometry with a total length of 90~m (referred to as Mont Blanc~\footnote{As the highest mountain in the Alps overall, Mont Blanc serves a the name for the configuration that corresponds to the geometry of the SHiP experiment~\cite{albanese_bdfship_2023} that has the best projected phase-space coverage of the currently approved fixed-target experiments. }) has the length scales of the planned SHiP experiment at CERN~\cite{albanese_bdfship_2023}. This geometry was chosen because SHiP so far has the best projected phase-space coverage. 

\begin{table}
    \centering
    \begin{tabular}{c | c | c}
        ~ & ~~Zugspitze~~ & ~~Mont Blanc~~ \\ \hline
        ~~$L_d$~~ & 2 m & 40 m \\ \hline
        ~~$L_v$~~ & 10 m & 50 m \\ \hline
        ~~$D_d$~~ & 10 m & 10 m \\
    \end{tabular}
    \caption{\it The two geometrical configuration that were considered in the simulations. }
    \label{tab:simulationGeometry}
\end{table}

\subsection{Compton Photon Source}

The phase-space coverage of a fixed-target experiment depends on the number of particles and their energy. Therefore, simulations were performed to find the optimal laser intensity for a specific electron beam energy that produces the highest number of photons with an energy of more than 1~GeV. A small polar angle of less than 2.5~mrad was required so that the photons can be used at the target in the forward direction from the electron-laser interaction point. 

The electron-laser interaction was simulated using Ptarmigan v1.4.0~\cite{blackburn_simulations_2023, blackburn_tgblackburnptarmigan_2024}. Ptarmigan is a Monte Carlo program that enables the simulation of electron-laser or photon-laser interaction in the transition regime from perturbative to non-perturbative QED where the local monochromatic approximation (LMA) is valid. However, it also allows simulations with the local constant field approximation (LCFA) that is used by modern particle-in-cell (PIC) codes, which is required and used here due to the high laser intensities. The simulation settings were a laser wavelength of $\lambda = 800~\mathrm{nm}$, a laser waist of $w_0 = 2~\mathrm{\mu m}$, a laser pulse duration of $\tau_\mathrm{fwhm} = 20~\mathrm{fs}$, a collision angle of $\alpha = 17.2\mathrm{\degree}$, a linear laser polarization, and a point-like electron beam. 

\begin{figure}
    \centering
    \includegraphics[width=1.0\linewidth]{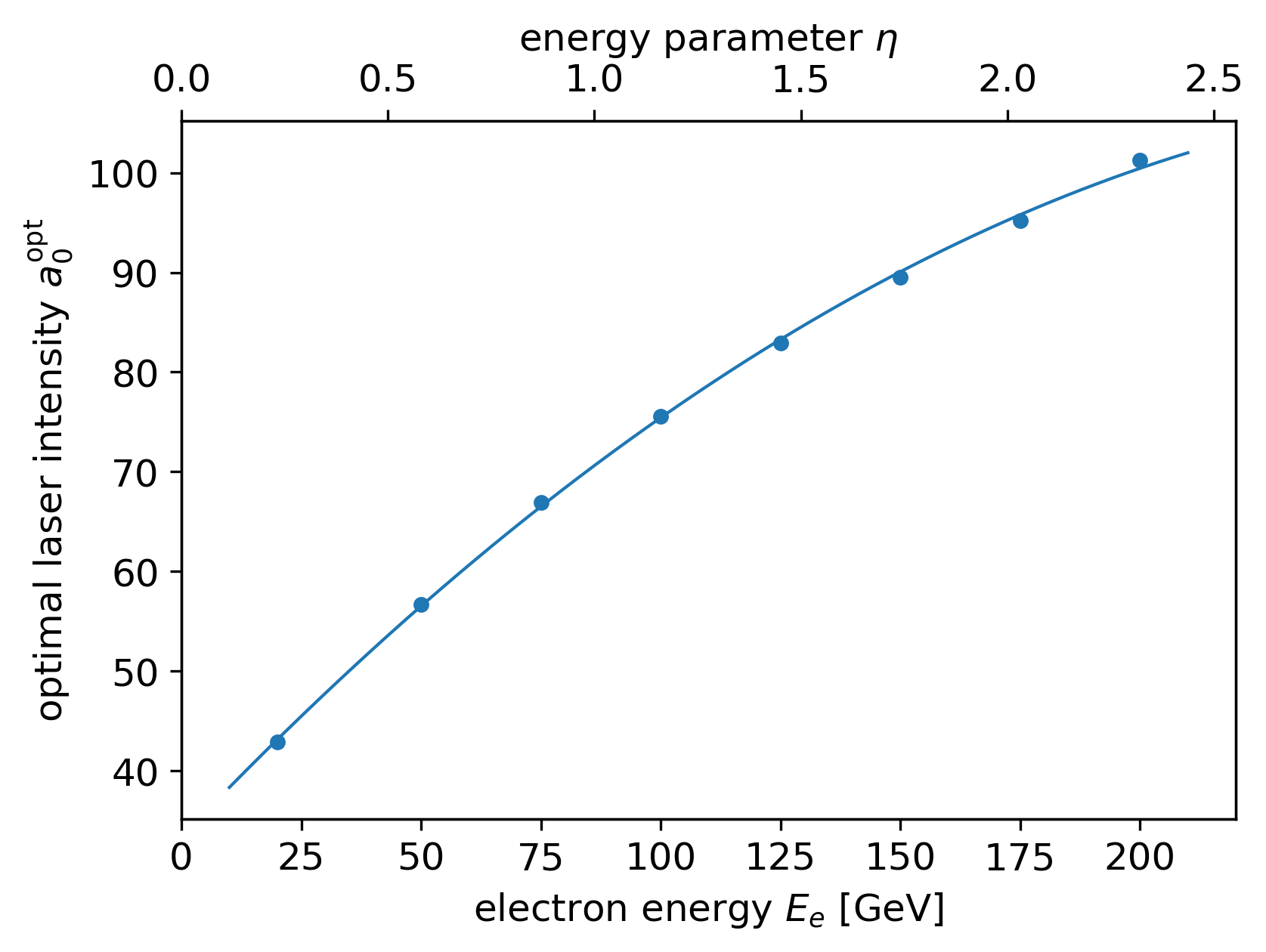}
    \caption{\it Combination of the laser intensity $a_0$ and the electron energy $E_e$ that produces the most high-energy Compton photons in the forward direction. The simulated points are fitted with a quadratic function. }
    \label{fig:bestA0}
\end{figure}

Figure~\ref{fig:bestA0} shows the combination of laser intensity and electron energy that produces the most high-energy photons in the forward direction. The electron energy of the European XFEL demands $a_0 \approx 40$ and the electron energy of 125~GeV at a future electron-positron collider a $a_0 \approx 80$. When examining existing laser systems such as Apollon (FR)~\cite{papadopoulos_apollon_2016}, ELI-NP (RO)~\cite{radier_10_2022}, or CoReLS (KR)~\cite{yoon_realization_2021}, they are capable of delivering laser intensities of $a_0 \gtrsim 200$, which is more than sufficient, but operate at low repetition rates below 1~Hz. Considering the progress of such systems~\cite{mourou_exawatt-zettawatt_2012, danson_petawatt_2019, khazanov_exawatt_2023}, it can be expected that the required power and rates will be available by the time a future electron facility begins to operate. 

At laser intensities of $a_0 \gtrsim 15$, the optical dump condition Eq.~(\ref{eq:opticalDumpCondition}) is partially violated because the time scale of the pair-prodution becomes shorter than the laser pulse length, i.e. ${t_l \gtrsim \tau_{ee}}$. This results in some Compton photons being converted into electron-positron pairs, which may be an additional source of background. Still, these laser settings yield the most high-energy photons. 

Various combinations of electron energy and laser intensity were simulated that correspond to possible future facilities. The most relevant parameters are summarized in Table~\ref{tab:facilityParameters} and discussed in more detail below. 

\begin{table}
    \centering
    \begin{tabular}{l | c | c | c | c}
        ~ & $E_e$ [GeV] & $N_e$ [$10^{10}$] & $R$ [Hz] & $a_0$ \\ \hline
         XFEL-LUXE & 16.5 & 0.15 & 1 & 3.4 \\ \hline
         XFEL-optimized & 17.5 & 0.62 & 10 & 41 \\ \hline
         FCCee-IC~& 45.6~\footnote{energy in the Z mode, which is the planned mode of operation for the first four years} & 21.7 & 1000 & 0.002 \\ \hline
         FCCee-Injector & 20 & 2.5 & 100~\footnote{the injector is only available for about 27\% of the time in the Z mode} & 43 \\ \hline
         ILC-like & 125 & 2 & 5 & 83 \\
    \end{tabular}
    \caption{\it Summary of the most relevant parameters of the electron-laser collisions that were simulated for the Compton photon sources. $E_e$ is the electron energy, $N_e$ the number of electrons per bunch, $R$ the repetition rate of the collisions, and $a_0$ the laser intensity parameter. }
    \label{tab:facilityParameters}
\end{table}

The scenario labeled XFEL-LUXE corresponds to the parameters considered in the LUXE-NPOD case at the European XFEL with a moderate laser intensity~\cite{bai_new_2022}. The configurations labeled XFEL-optimized and ILC-like correspond to the parameters of the European XFEL and a facility like the International Linear Collider~\cite{aryshev_international_2023}, respectively, with an optimized laser intensity as found in the simulations presented in Fig.~\ref{fig:bestA0}. The case for a circular collider like the Future Circular Collider (FCCee) at CERN is more complex, since bunches cannot be continuously extracted. There are various options: 
\begin{itemize}
    \item The intensity control of the collider applies Compton backscattering to balance the population of the colliding bunches~\cite{drebot_optimizing_2023}. It has the Compton source with the highest laser intensity included in the \mbox{FCCee} design and is labeled FCCee-IC here. 
    \item The FCCee injector complex, labeled FCCee-Injector, is only used part-time to top up the collider and can be used between times for the beyond-collider physics program~\cite{bartosik_possibilities_2024}. 
    \item In principle, also the FCCee booster is available for some time in the high-energy modes (i.e. WW, Z(HZ), and ttbar)~\cite{bartosik_possibilities_2024}. However, since only single bunches can be boosted and collided with a laser, the rates are low, and this case is not considered in our study. 
    \item The EPOL, used for the polarization and energy calibration, applies Compton backscattering at low laser intensity~\cite{blondel_polarization_2019}. It is not shown here since it is not competitive due to the low laser intensity of ${a_0 \approx 4 \times 10^{-6}}$. 
\end{itemize}

There are other facilities that are not directly evaluated. The Circular Electron Positron Collider (CEPC)~\cite{the_cepc_study_group_cepc_2023} will be very similar to FCCee and comparable searches can be implemented. Other linear collider facilities like the Compact Linear Collider (CLIC)~\cite{brunner_clic_2022} or the Cool Copper Collider (C$^3$)~\cite{dasu_strategy_2022} provide a bunch structure different from ILC which increases the luminosity by a factor of 2.6 or 7.4, respectively. This is because these facilities provide the electrons in a bunch-train structure and high-power laser systems can only trigger on a single bunch per train. Therefore, linear accelerator facilities with a high train rate are better suited, since the higher luminosity extends the phase-space coverage in all dimensions, as shown in Ref.~\cite{bai_new_2022}. The most recent linear collider concept is the Hybrid, Asymmetric, Linear Higgs Factory (HALHF)~\cite{foster_hybrid_2023}. This facility would provide even more electrons than any of the other future linear machines at a higher energy of 500~GeV for the same center-of-mass energy of the collisions. For this study, we decided to investigate in detail only a set of facilities that could be realized in the near future. 

\begin{figure}
    \centering
    \includegraphics[width=1.0\linewidth]{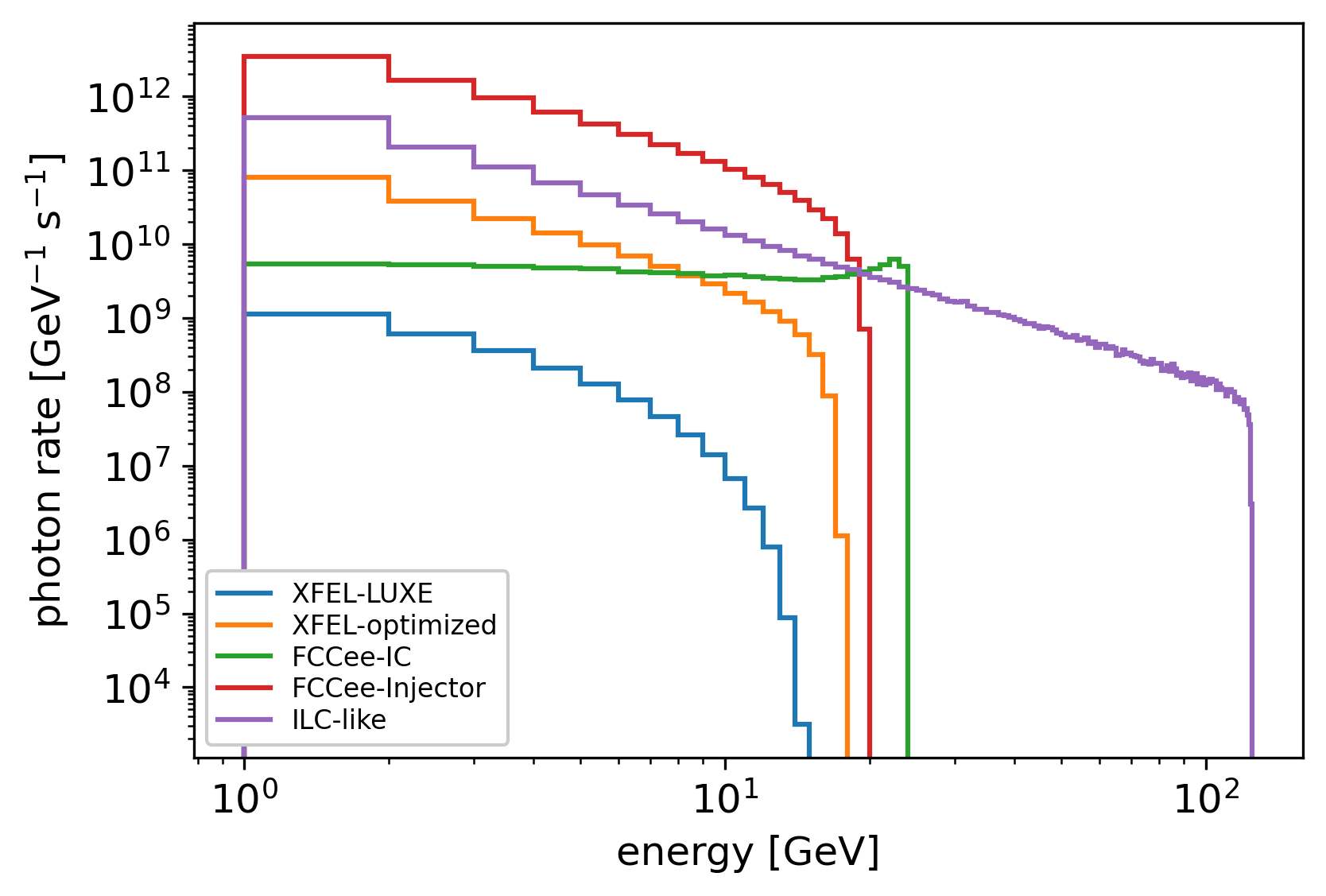}
    \caption{\it Compton photon spectra for the different settings of the electron-laser collisions given in Table~\ref{tab:facilityParameters}. }
    \label{fig:comptonPhotonSpectra}
\end{figure}

For the above-mentioned configurations, the (non-linear) Compton process was simulated using Ptarmigan. The energy spectrum of the FCCee intensity control was taken from Ref.~\cite{drebot_optimizing_2023} where the simulations were performed with CAIN~\cite{chen_cain_1995}. The Compton photon spectra are shown in Fig.~\ref{fig:comptonPhotonSpectra}. 

\section{Background}\label{sec:background}

Searches for new particles beyond the Standard Model have to be sensitive to a handful of events, which requires a background-free environment. This can be achieved by suppressing the number of background events with thick shielding, by a good particle identification, and by a high rejection power for those background particles that still reach the detector. In the case of LUXE-NPOD, extensive studies and simulations have been performed~\cite{bai_new_2022, trevisani_probing_2024}. It was found that a background-free environment ($N_\mathrm{bkg} \lesssim 1$ per year) can be achieved with a tungsten dump of $L_d = 1~\mathrm{m}$ with an additional lead wrap and a concrete wall and detector technology that is currently available. The average photon energy in that scenario for Compton photons above 1~GeV is ${E_c^\mathrm{avg}= 3.1~\mathrm{GeV}}$. 

The first background estimates have been done for ILC-NPOD for the ILCX workshop in 2021~\cite{soreq_probing_2021}. The number of background events has roughly an exponential behaviour
\begin{equation}\label{eq:backgroundScaling}
    N_\mathrm{bkg} \propto N_c E_c^\mathrm{avg} \, e^{-L_d / L_\mathrm{nuc}} \ .
\end{equation}
In the case of tungsten $L_\mathrm{nuc} \approx (5 - 10)~\mathrm{cm}$, where the relevant length scales are the nuclear collision length ${\lambda_T = 5.719~\mathrm{cm}}$ and the nuclear interaction length ${\lambda_I = 9.946~\mathrm{cm}}$~\cite{navas_review_2024}. The average Compton photon energy in this case was ${E_c^\mathrm{avg}= 39~\mathrm{GeV}}$ and the number of Compton photons $N_c$ per colliding electron was comparable. Taking into account the higher photon energy (by a factor of 12.6) and the higher number of colliding electrons (by a factor of 66.7), it was estimated using Eq.~(\ref{eq:backgroundScaling}) that ILC-NPOD should also be background-free if the dump length is increased to $L_d \simeq 2~\mathrm{m}$. 

This estimate holds true only when the spectra are comparable, allowing the use of the average photon energy. This is not the case here, since the simulations cover very different combinations of electron energies and laser intensities, as shown in Table~\ref{tab:facilityParameters}. To account for this, the product $ N_c E_c^\mathrm{avg}$ has to be replaced by the integral over the energy spectrum and the required dump length can be calculated using 
\begin{equation}\label{eq:dumpLengthRequirement}
    L_d^\mathrm{fc} = L_d^\mathrm{luxe} - \ln \left( \frac{N_e^\mathrm{luxe} R^\mathrm{luxe} \int N_c^\mathrm{luxe} ( E ) \, \mathrm{d}E}{N_e^\mathrm{fc} R^\mathrm{fc} \int N_c^\mathrm{fc} ( E ) \, \mathrm{d}E} \right) L_\mathrm{nuc} \ ,
\end{equation}
where the superscript '$\mathrm{luxe}$' represents the parameters of the LUXE-NPOD study that showed the background-free environment and the superscript '$\mathrm{fc}$' represents any of the future collider facilities, and $R$ is the repetition rate of the electron-laser collisions. Table~\ref{tab:dumpLengths} summarizes the results for the different settings. In particular, the configuration using the FCCee injector requires the longest dump with a minimum length of 1.83~m, even though the photon energies are much lower. The calculations confirm the estimates Ref.~\cite{soreq_probing_2021} and show that a dump with a length of $L_d = 2~\mathrm{m}$ as in the Zugspitze geometry is sufficient to achieve a backgroud-free scenario for all examined photon sources. 

\begin{table}
    \centering
    \begin{tabular}{l | c}
         &$L_d$  \\ \hline
         XFEL-optimized & ~1.45~m~ \\ \hline
         FCCee-IC~ & ~1.51~m~\\ \hline
         FCCee-Injector & ~1.83~m~\\ \hline
         ILC-like & ~1.68~m~\\
    \end{tabular}
    \caption{\it Minimum length of the dump $L_d$ for the different settings that result in a background-free environment. }
    \label{tab:dumpLengths}
\end{table}

\section{Results}\label{sec:results}

\begin{figure}
    \centering
    \includegraphics[width=1.0\linewidth]{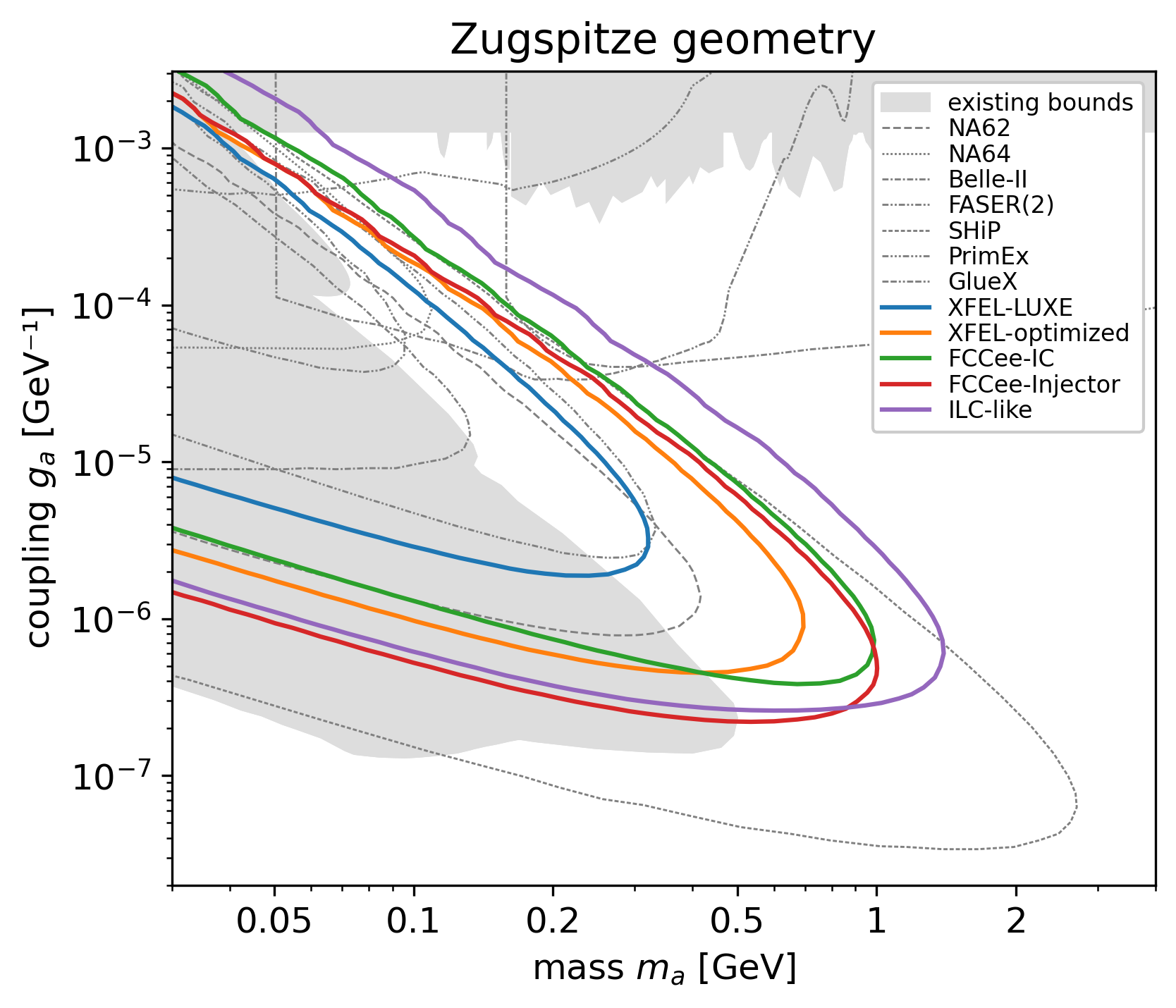}
    \caption{\it Projected phase-space coverage of various experimental settings for the coupling of ALPs to photons. The coverage was calculated for the Zugspitze geometry defined in Table~\ref{tab:simulationGeometry}. Same color code as in Fig.~\ref{fig:comptonPhotonSpectra}. The gray shaded areas correspond to existing bounds from laboratory experiments~\cite{bergsma_search_1985, riordan_search_1987, bjorken_search_1988, blumlein_limits_1991, dolan_revised_2017, knapen_searching_2017, aloni_photoproduction_2019, dobrich_light_2019, banerjee_search_2020, abudinen_search_2020, ablikim_search_2023, capozzi_new_2023, faser_collaboration_shining_2024}. Projections for current experiments are indicated by various dashed lines~\cite{dolan_revised_2017, feng_axionlike_2018, dobrich_light_2019, aloni_photoproduction_2019, dusaev_photoproduction_2020, albanese_bdfship_2023}. }
    \label{fig:phaseSpace_benchmark}
\end{figure}

\begin{figure}
    \centering
    \includegraphics[width=1.0\linewidth]{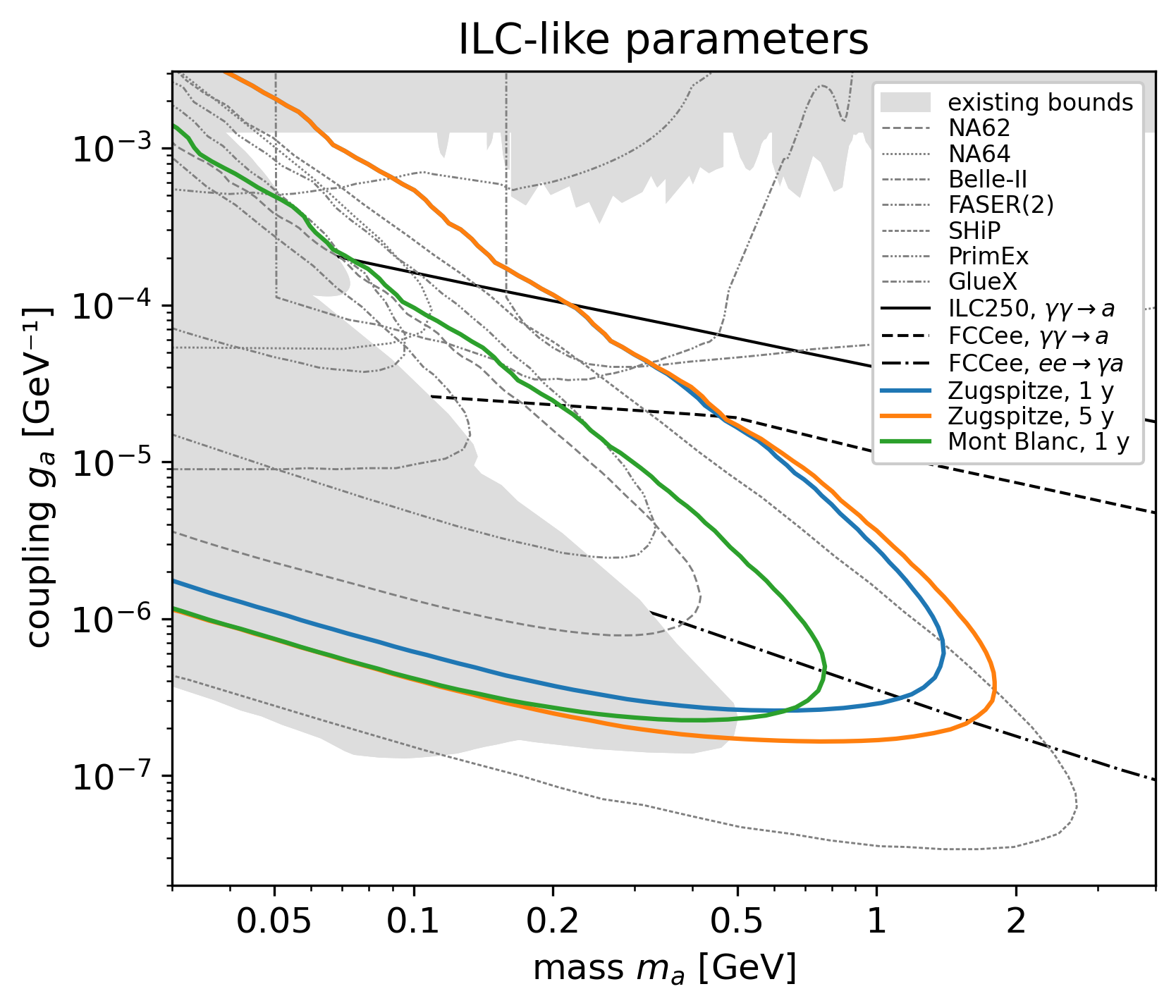}
    \caption{\it Same as ILC-like coverage in Fig.~\ref{fig:phaseSpace_benchmark} (blue), for a runtime of 5 years (orange), as well as the Mont Blanc geometry defined in Table~\ref{tab:simulationGeometry} (green). Projections from direct searches of ILC250 and FCCee via $\gamma\gamma \to a$~\cite{teles_searches_2024} and for FCCee via $e^+ e^- \to \gamma a$~\cite{bauer_axion-like_2019} are also shown. }
    \label{fig:phaseSpace_ilcLikeComparisons}
\end{figure}

With the Compton-photon spectra from the different electron-laser collision settings and the calculated Primakoff production cross section, we simulated the production of the ALPs and their subsequent decay into two photons. For each combination of ALP mass $m_a$ and coupling to photons $g_a$, we counted the number of particles that met the conditions described in Sec.~\ref{sec:geometry}. Assuming a runtime of $10^7$~s, corresponding to 1~year with 32\% uptime, and a background-free scenario, the three-event boundary marks the phase-space coverage on a 95\% confidence limit. These boundaries for the Zugspitze geometry are presented in comparison with existing laboratory constraints~\cite{bergsma_search_1985, riordan_search_1987, bjorken_search_1988, blumlein_limits_1991, dolan_revised_2017, knapen_searching_2017, aloni_photoproduction_2019, dobrich_light_2019, banerjee_search_2020, abudinen_search_2020, ablikim_search_2023, capozzi_new_2023, faser_collaboration_shining_2024} and projections of current experiments~\cite{dolan_revised_2017, feng_axionlike_2018, dobrich_light_2019, aloni_photoproduction_2019, dusaev_photoproduction_2020, albanese_bdfship_2023} in Fig.~\ref{fig:phaseSpace_benchmark}. Figure~\ref{fig:phaseSpace_ilcLikeComparisons} compares the ILC-like scenario for different geometries and runtimes, as well as to direct searches of future colliders. 

Figure~\ref{fig:phaseSpace_benchmark} shows that all simulated projections cover parts of the ALP-photon coupling phase-space that is yet unconstraint. It also shows that a higher beam energy allows access to heavier ALP masses. This is due to the higher production cross section and the longer decay length due to the higher momentum that allows them to leave the dump volume before decaying. Figure~\ref{fig:phaseSpace_ilcLikeComparisons} clearly shows the advantage of having a shorter dump as in the Zugspitze configuration. 

Comparing the estimated phase-space coverage of the simulated configurations with the latest SHiP projection for a runtime of 15~years~\cite{albanese_bdfship_2023} indicates that there is a relevant new physics reach in the scenario of a linear collider facility such as the ILC with a beam energy of 125~GeV. This scenario should be further optimized and investigated. Moreover, such a facility would possibly allow for energy upgrades in the future which would allow access to heavier masses. However, projections for direct searches at future colliders reveal excellent sensitivity. Especially a circular collider like the FCCee with an outstanding luminosity shows great detection capability via the process $e^+ e^- \to \gamma a$. 

Although our results indicate that no new phase-space can be explored beyond the projections of the SHiP experiment and direct searches at future collider facilities, our approach remains highly complementary. Unlike existing fixed-target experiments that rely on charged particles, our method leverages photon interactions via the Primakoff mechanism directly. This offers a distinct and independent probe of the ALP-photon coupling. Given that the exact parameter space where ALPs may exist remains unknown, a diverse range of search strategies is essential to maximize discovery potential. Future refinements in experimental design and photon beam intensities could further enhance the sensitivity of this approach, reinforcing the importance of exploring multiple avenues in the search for physics beyond the Standard Model.

\section{Conclusion}\label{sec:conclusion}

In conclusion, we showed how future lepton colliders and laser infrastructures could be utilized to search for new particles beyond the Standard Model of particle physics by exploiting strong-field QED experimental setups. The photons produced in such experiments, but also at other sections of the colliders, via the Compton process, can be further used to search for new physics in fixed-target experiments. So far, no such search has been performed that uses photons as primary particles. Our analysis revealed which part of the phase-space of ALPs coupling to photons can be covered for two different experimental geometries and collider and laser settings. For more realistic results, the following should be considered in future studies.

\begin{itemize}
    \item To confirm the estimates of the background-free scenario described in Sec.~\ref{sec:background}, thorough studies must be performed. 
    \item In this study, only primary photons were considered for Primakoff production. In the LUXE configuration, this is valid because the secondary photons have low energy and do not contribute to the signal. However, at a future collider with beam energy of 125~GeV, this does not have to be the case and should be investigated. 
    \item We assumed a perfect point-like electron beam with no divergence which should be addressed for more realistic results. 
    \item For the case of LUXE-NPOD, it was shown that the optical dump has a highly suppressed background compared to the case where electrons are dumped directly~\cite{bai_new_2022}. However, this might be different in other configurations and requires more detailed studies. 
\end{itemize}

\begin{acknowledgments}
We gratefully acknowledge all the fruitful discussions with I.~Drebot on Compton backscattering, the intensity control of FCCee, and for providing its spectral data. This work was supported by the Swiss National Science Foundation under grant No. 214492. 
\end{acknowledgments}

\section*{Conflict of Interest}
The authors have no conflicts of interest to disclose. I.~Schulthess is involved in the beyond-collider physics studies of future linear and circular collider facilities. F.~Meloni is not involved in any of the future collider projects considered.

\section*{Author Contributions}
\textbf{I.~Schulthess:} conceptualization (equal); data curation (lead); formal analysis (lead); funding acquisition (lead); investigation (equal); methodology (equal); software (lead); validation (equal); visualization (lead); writing - original draft (lead); writing – review \& editing (equal). 
\textbf{F.~Meloni:} conceptualization (equal); investigation (supporting); methodology (equal); validation (equal); writing – review \& editing (equal). 

\section*{Data Availability}
The data and analysis that support the findings of this study are openly available in a Github repository~\cite{schulthess_ivoschulthessfuturecollider_npod_2025}.

\bibliography{references_bibtex.bib}

\end{document}